\documentclass[sigconf,authorversion]{acmart}

\usepackage{caption}
\usepackage{subcaption}

\usepackage{placeins} % -> use \FloatBarrier

\usepackage{stfloats}
\usepackage{float}

\usepackage[colorinlistoftodos,prependcaption,disable]{todonotes} %put in disable to make them invisible
\presetkeys%
    {todonotes}%
    {inline,backgroundcolor=orange}{}
    
\usepackage[nopostdot,acronym,shortcuts,nonumberlist]{glossaries}
\newacronym{IoT}{IoT}{Internet of Things}
\newacronym{BLE}{BLE}{Bluetooth Low Energy}
\newacronym{GPS}{GPS}{Global Positioning System}
\newacronym{CERT}{CERT}{Computer Emergency Response Team}
\newacronym{API}{API}{Application Programming Interface}
\newacronym{MITM}{MITM}{Machine-in-the-Middle}
\newacronym{TLS}{TLS}{Transport Layer Security}
\newacronym{OTA}{OTA}{Over-the-Air}
\newacronym{XSS}{XSS}{Cross-Site Scripting}
\newacronym{SDK}{SDK}{Software Development Kit}
\newacronym{MQTT}{MQTT}{Message Queuing Telemetry Transport}
\newacronym{CVE}{CVE}{Common Vulnerabilities and Exposures}
\newacronym{ECDH}{ECDH}{Elliptic-curve Diffie–Hellman}
\newacronym{AEAD}{AEAD}{Authenticated Encryption with Associated Data}
\newacronym{AES}{AES}{Advanced Encryption Standard}
\newacronym{TOR}{TOR}{The Onion Router}
\newacronym{GDPR}{GDPR}{EU General Data Protection Regulation}
\newacronym{HSTS}{HSTS}{HTTP Strict Transport Security}
\newacronym{AWS}{AWS}{Amazon Web Services}
\newacronym{RPA}{RPA}{Resolvable Private Address}
\newacronym{IRK}{IRK}{Identity-Resolving Key}
\newacronym{DoS}{DoS}{Denial of Service}

\usepackage{pgf-umlsd-customized}
\usetikzlibrary{calc}
\tikzset{
    >=triangle 45
}
\usepackage{siunitx}

\usepackage{tabularx}
\usepackage{slashbox}

\definecolor{sorange}{rgb}{0.95, 0.57, 0}
\colorlet{orange}{sorange}

\usepackage{listings}
\usepackage{xspace}

\acmYear{2020}\copyrightyear{2020}
\setcopyright{acmcopyright}
\acmConference[WiSec '20]{13th ACM Conference on Security and Privacy in Wireless and Mobile Networks}{July 8--10, 2020}{Linz (Virtual Event), Austria}
\acmBooktitle{13th ACM Conference on Security and Privacy in Wireless and Mobile Networks (WiSec '20), July 8--10, 2020, Linz (Virtual Event), Austria}
\acmPrice{15.00}
\acmDOI{10.1145/3395351.3399422}
\acmISBN{978-1-4503-8006-5/20/07}

\acmSubmissionID{96}

\begin{document}

\title{Lost and Found: Stopping Bluetooth Finders from \\ Leaking Private Information}


\author{Mira Weller}
\affiliation{%
  \institution{Secure Mobile Networking Lab}
  \institution{TU Darmstadt, Germany}
%  \streetaddress{Pankratiusstraße 2}
%  \city{Darmstadt}
%  \country{Germany}
%  \postcode{64293}
}
\email{mweller@seemoo.de}

\author{Jiska Classen}
\affiliation{%
  \institution{Secure Mobile Networking Lab}
  \institution{TU Darmstadt, Germany}
%  \streetaddress{Pankratiusstraße 2}
%  \city{Darmstadt}
%  \country{Germany}
%  \postcode{64293}
}
\email{jclassen@seemoo.de}

\author{Fabian Ullrich}
\affiliation{%
  \institution{Secure Mobile Networking Lab}
  \institution{TU Darmstadt, Germany}
%  \streetaddress{Pankratiusstraße 2}
%  \city{Darmstadt}
%  \country{Germany}
%  \postcode{64293}
}
\email{fullrich@seemoo.de}

\author{Denis Wa{\ss}mann}
\affiliation{%
  \institution{Secure Mobile Networking Lab}
  \institution{TU Darmstadt, Germany}
%  \streetaddress{Pankratiusstraße 2}
%  \city{Darmstadt}
%  \country{Germany}
%  \postcode{64293}
}
\email{dwassmann@seemoo.de}

\author{Erik Tews}
\affiliation{%
  \institution{EEMCS Department, SCS Group}
  \institution{University of Twente, Netherlands}
  %\streetaddress{TODO}
%  \city{Twente}
%  \country{Netherlands}
}
\email{e.tews@utwente.nl}

%\author{Matthias Hollick}
%\affiliation{%
%  \institution{Secure Mobile Networking Lab, TU Darmstadt}
%  \streetaddress{Mornewegstrasse 32}
%  \city{Darmstadt}
%  \state{Hesse}
%  \postcode{64293}
%}

\renewcommand{\shortauthors}{Weller et al.}
\renewcommand{\shorttitle}{Lost and Found: \emph{PrivateFind}}
\newcommand{\mytool}{{\emph{PrivateFind}}\xspace} %TODO name

%!TEX root = ../finder.tex

% The abstract is a short summary of the work to be presented in the article.
\begin{abstract}
A Bluetooth finder is a small battery-powered device that can be attached to important items such as bags, keychains, or bikes. The finder maintains a Bluetooth connection with the user's phone, and the user is notified immediately on connection loss.
%Many products
%include all enrolled users in searching for anyone's lost Bluetooth finder and report their locations to the legitimate owners through a cloud service. %Maybe split this sentence, it is very confusing. %redundant anyway...
We provide the first comprehensive security and privacy analysis of current commercial Bluetooth finders. 
Our analysis reveals several significant security vulnerabilities in those products concerning mobile applications and the corresponding backend services in the cloud. We also show that all analyzed cloud-based products leak more private data than required for their respective cloud services.

Overall, there is a big market for Bluetooth finders, but none of the existing products is privacy-friendly.
We close this gap by designing and implementing \mytool, which ensures
locations of the user are never leaked to third parties. It is designed to run on similar hardware as existing finders, allowing vendors to update their systems using \mytool.
\end{abstract}




\maketitle

\newpage %authorvers fix
%!TEX root = ../finder.tex

\section{Introduction}

Bluetooth finders are small devices that track an item's location.
The finder's owner installs a smartphone app that maintains a \ac{BLE}
connection with the finder. If the connection is interrupted, the app assumes that the corresponding item
is lost and alerts the owner---for example, when leaving the house without the wallet.
The limited \ac{BLE} range allows triggering alarms accurately. In case a user is attempting to locate a nearby misplaced item, the app is able to play a sound on a given finder within wireless range.
Localization also works in the reverse direction: Upon pressing the finder's button the smartphone plays a sound.

In addition to the sound-based localization, the app saves the smartphone's \ac{GPS} location and associates it with
the finder. Position accuracy depends on the \ac{GPS} fix and the maximum distance
between finder and smartphone. The owner is then able to retrieve the last known location of the finder. 
Position data is saved locally or on a remote server.

\begin{figure*}[!t]
	\begin{center}
	\begin{tikzpicture}[minimum height=0.55cm, scale=0.8, every node/.style={scale=0.8}, node distance=0.7cm]

    \node[inner sep=0pt] (app) at (-7,-0.2)
    {\includegraphics[width=1cm]{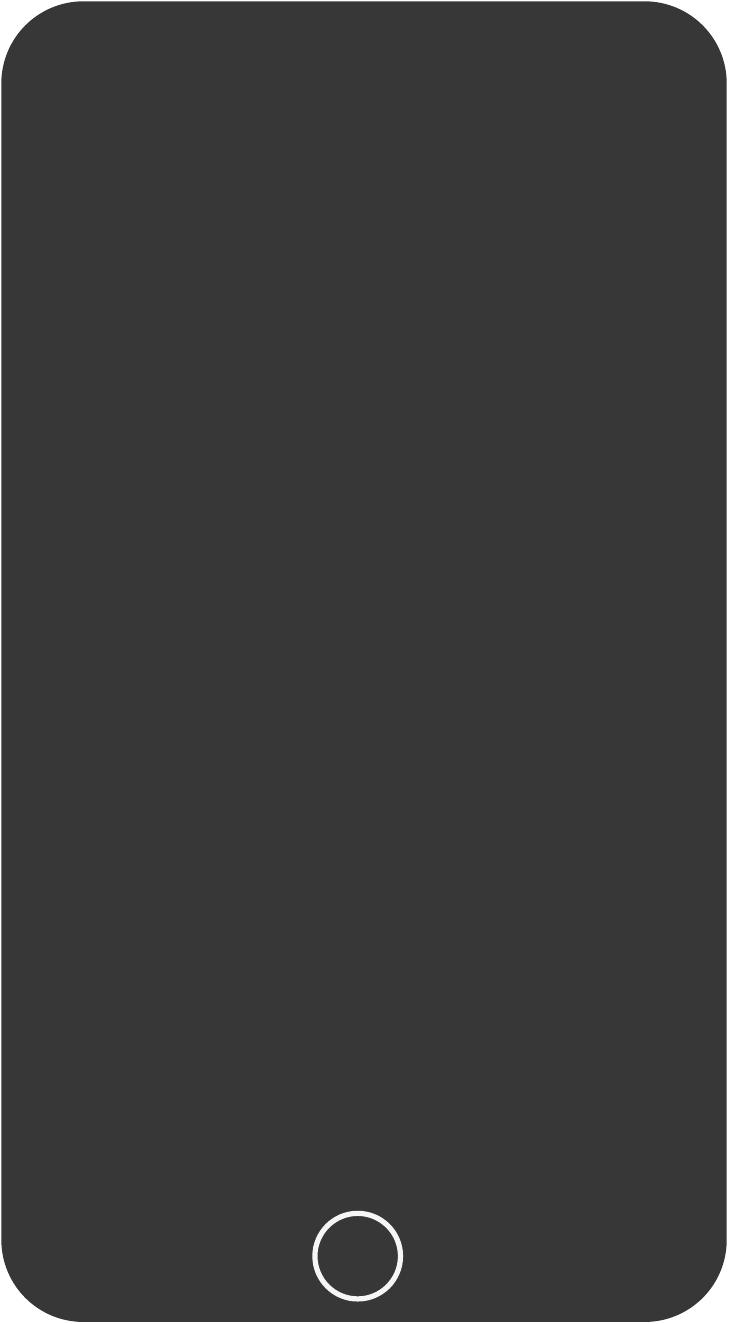}};
    \node[inner sep=0pt] (screenshot) at (-7,-0.2)
    {\includegraphics[width=0.8cm]{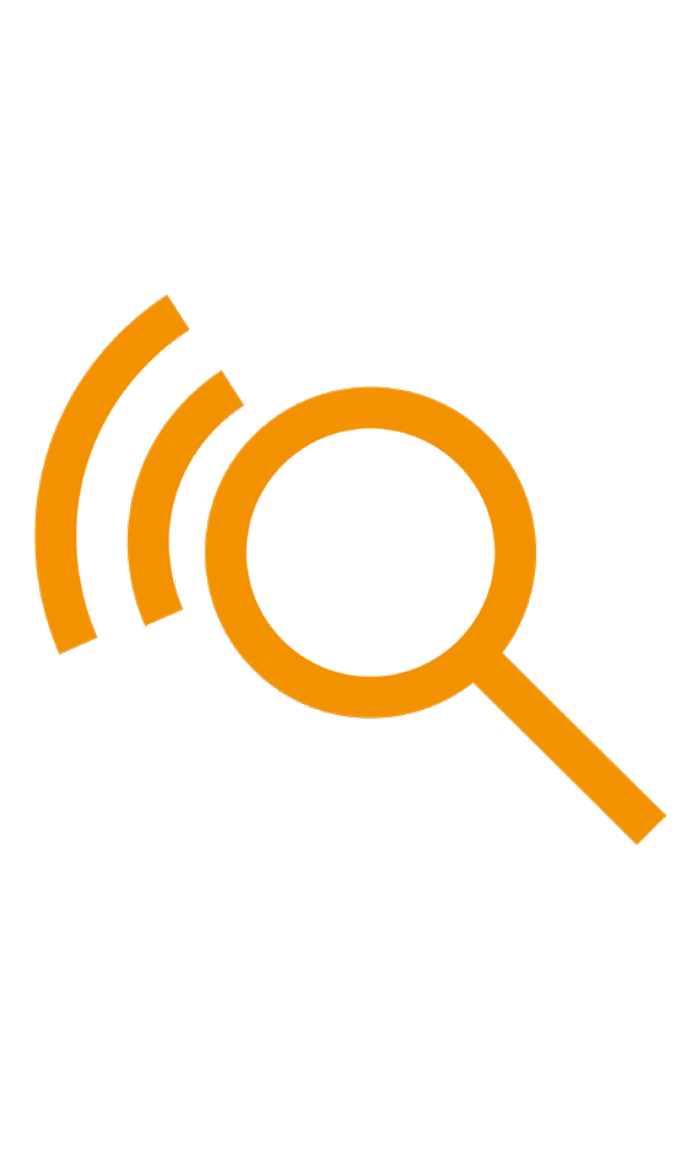}};
    \node[inner sep=0pt] (app2) at (-1.7,0.7)
    {\includegraphics[width=1cm]{pics/smartphone.pdf}};
    \node[inner sep=0pt] (screenshot2) at (-1.7,0.7)
    {\includegraphics[width=0.8cm]{pics/app.png}};
    \node[inner sep=0pt] (server) at (-13,0.5)
    {\includegraphics[width=2.7cm]{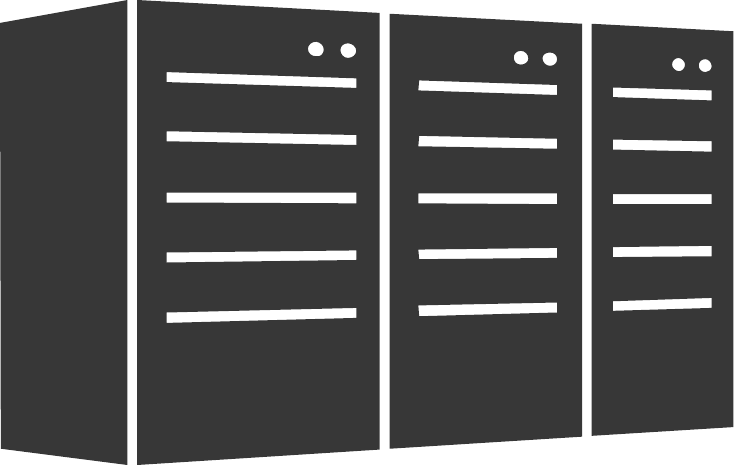}};
    \node[inner sep=0pt] (gps) at (-5.8,3.4)
    {\includegraphics[width=2.5cm,angle=30]{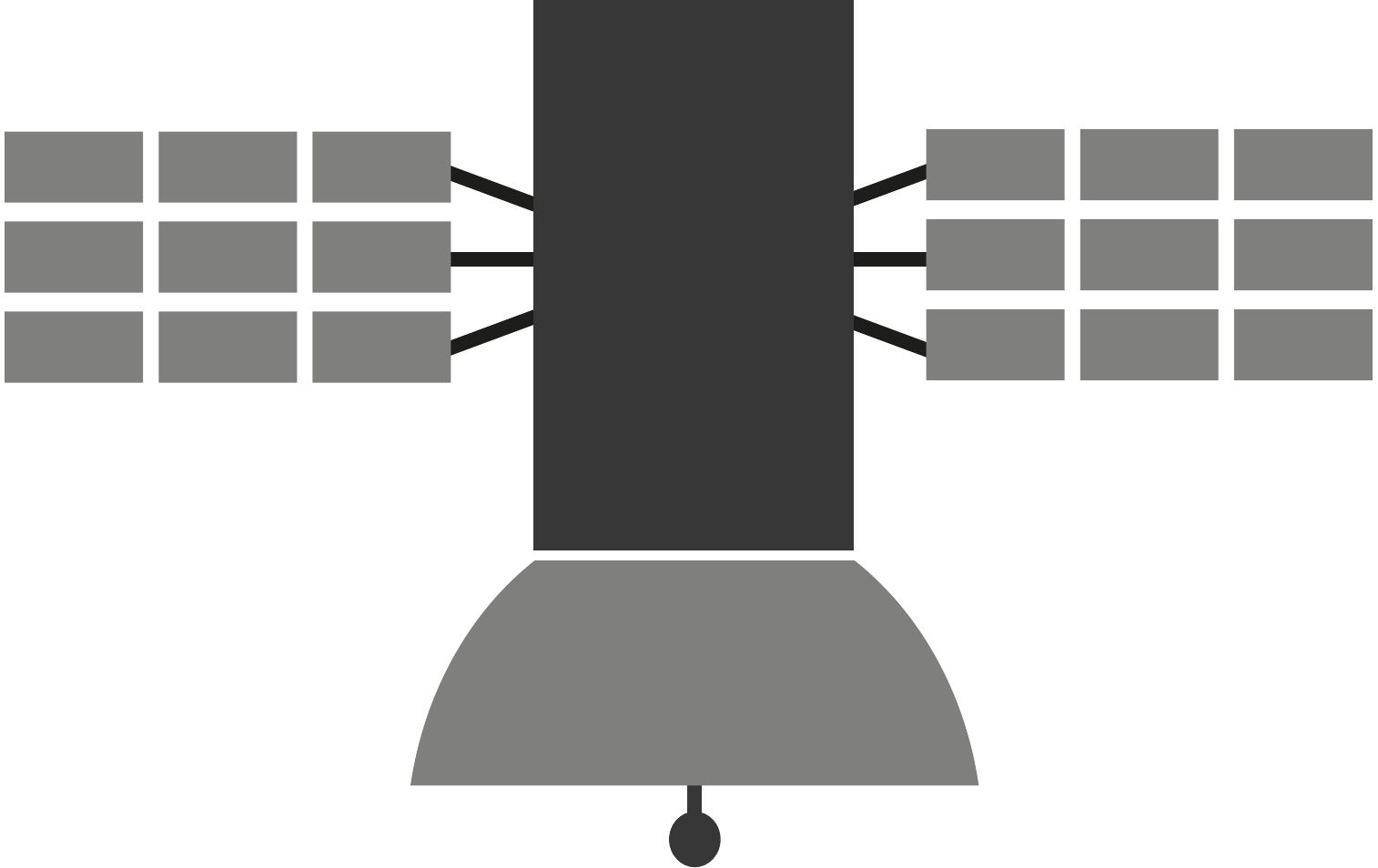}};
    \node[inner sep=0pt] (gps2) at (0.2,3.3)
    {\includegraphics[width=2.5cm,angle=-20]{pics/sat.pdf}};
    \node[inner sep=0pt] (cat) at (2,0)
    {\includegraphics[width=2.5cm]{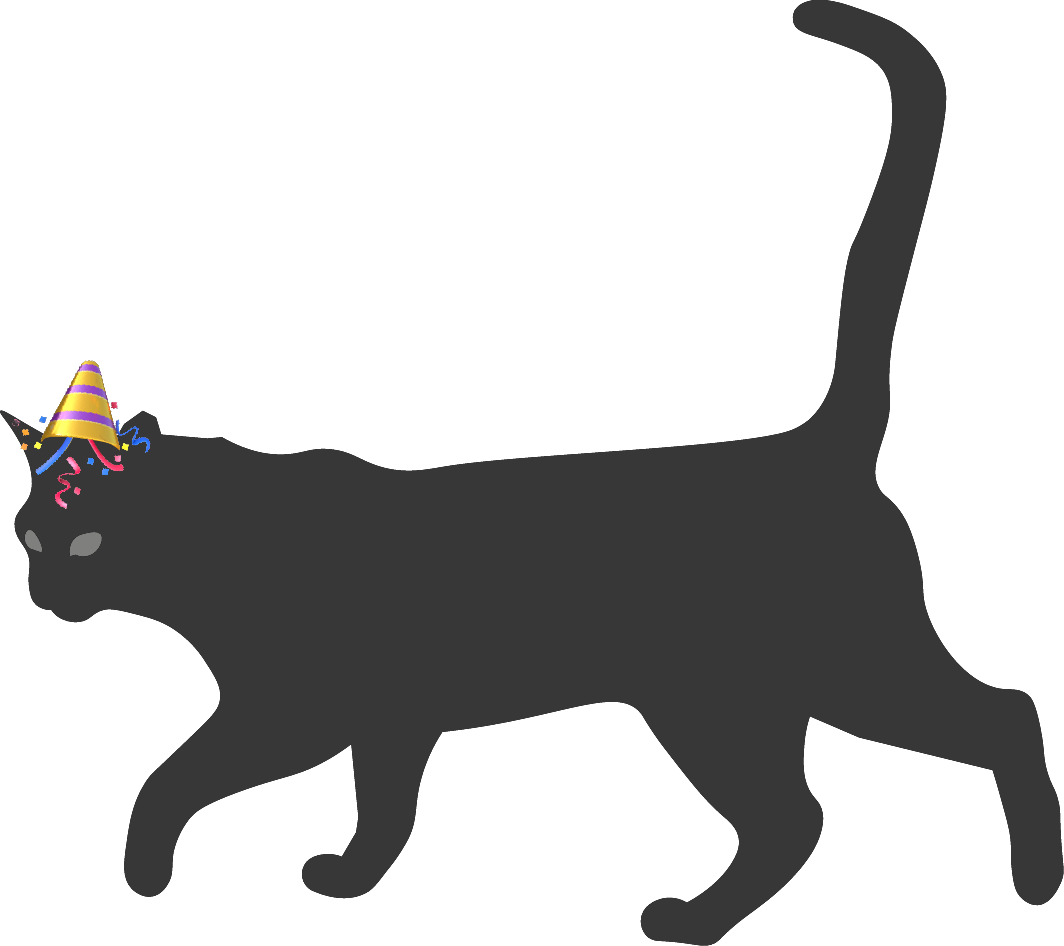}};
    \node[inner sep=0pt] (finder) at (2.6,0.7)
    {\includegraphics[width=0.5cm,angle=10]{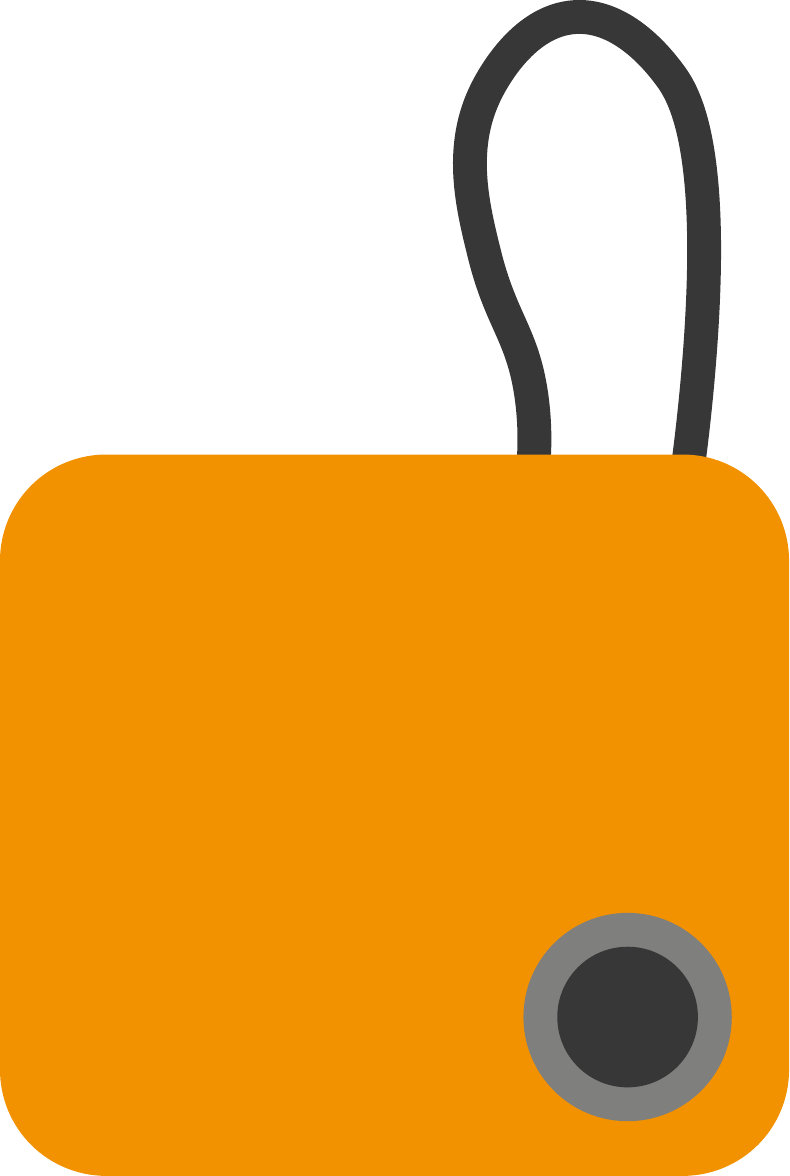}};

    \node[above=of app.south, anchor=north, yshift=-0.8cm] (apptxt) {Owner};
    \node[above=of app2.south, anchor=north, yshift=-0.8cm] (app2txt) {Reporter};
    \node[above=of finder.north, anchor=north, yshift=-0.3cm] (findertxt) {Finder};
    \node[above=of server.north, anchor=north, yshift=-0.3cm] (servertxt) {Server};

	\path[<->,dotted,thick,color=orange] (-1.1,0.7) edge node[sloped, anchor=center, above, text width=2.3cm, font=\footnotesize] {Finder information} (finder.west);
	\path[<-,dotted,thick,color=gray] (-2.3,1.6) edge node[anchor=center, above, text width=3cm, font=\footnotesize] {GPS} (-5.3,2.7);
	\path[<-,dotted,thick,color=gray] (-1.1,1.6) edge node[anchor=center, above, text width=1cm, font=\footnotesize] {GPS} (-0.1,2.6);
	\path[<-] (-11.5,1) edge node[anchor=center, above, text width=3cm, font=\footnotesize] {Lost finder report} (-2.3, 1);
	\path[->] (-11.5,0) edge node[anchor=center, above, text width=2.5cm, font=\footnotesize] {Found your finder} (-7.7, 0);
	\end{tikzpicture}
	\end{center}
	\vspace{-1em} %TODO because the next section really breaks weird without this
\caption{Bluetooth finder ecosystem overview. The finder's owner lost connectivity, but the moving finder is no longer at its last known position. A reporter within Bluetooth range sends their GPS position to the owner via the server.}
\label{fig:overview}
\end{figure*}

A tracked item can move while it is out of its owner's Bluetooth range. For that reason, some finders allow other users to search for a lost
item.
All apps are continuously scanning for
currently disconnected finders in the background and report these to the server.
Once a lost finder is spotted by another user, the owner is informed.
\autoref{fig:overview} depicts the ecosystem architecture that is required
to support external position reporting and searching.

Bluetooth finders became popular in 2013 when \emph{Tile} raised \$\,2.6 million with a crowdfunding campaign~\cite{tilefund}.
Since then, various finders were introduced to the market.
In this work, we analyze top-ranked finders sold worldwide: \emph{Nut Find3}, \emph{Tile Mate}, \emph{Tile Pro}, \emph{musegear finder}, \emph{Pearl Callstel Key Finder}, \emph{Gigaset g-tag}, and \emph{Cube Tracker}. Additionally, we analyze a couple of lesser known brands based on the \emph{ST17H26} chip that can be managed over various smartphone apps including \emph{iTracing}, \emph{iSearching}, and \emph{FindELFI}. %This sentence is way to long. Maybe omit some of the tastes (Mate/Pro etc.) 
%\todo{\emph{iTrackEasy}, \emph{iTracing}, \emph{iSearching}, and \emph{FindELFI} are not names of finders but of the apps. the finders itself are noname}
% on platforms like \emph{Amazon}, \emph{eBay} and \emph{AliExpress}.
 Our analysis uncovers severe security and privacy issues.
We reported all issues to the vendors, and most have been fixed by now.  
However, despite being one of the top-selling finders, \emph{Nut} still has major security issues.
We communicated these more than a year ago over various communication channels and contacted two \glspl{CERT}, but
\emph{Nut} servers are still leaking user data.
%Their server side implementation
%leaks all lost tracker reporting data, which includes account data of the
%reporter, tracker positions and IDs, and valid mail addresses and phone numbers.
%We contacted them over various channels---including letter post, writing to
%the Chinese developers, social media direct messages and two \glspl{CERT}---but
%the issues remain and private data is accessible since months.
This demonstrates how even today adequate product security is still being neglected by some low-cost \ac{IoT} vendors.

Despite the variety of available Bluetooth finders, a solution maintaining privacy has yet to be proposed.
Therefore, we design and implement the privacy-preserving secure finder protocol \mytool for an \ac{IoT} platform similar to those seen on low-budget finders.
Our contributions are as follows:
\begin{itemize}
\item A comprehensive analysis of features, security, and privacy in popular Bluetooth finder ecosystems.
\item Uncovering severe security and privacy breaches in the most popular finders \emph{Tile} and \emph{Nut}.
\item Generalized findings of common design flaws in \ac{IoT} ecosystems that compromise security and privacy.
\item Design of a new solution \mytool supporting all observed features but providing anonymity and privacy.
\item Implementation of \mytool as an open-source project on a low-cost \gls{IoT} platform similar to existing finder hardware.
\end{itemize}
% The introduction has a lot of "we". Especially the above listing is problematic. Maybe more often chooe a passive form?

This paper is structured as follows.
We list previous security research in \autoref{sec:related}.
In \autoref{sec:analysis}, we analyze popular finder features, with a focus on their infrastructure and security.
We solve the identified issues while providing all important features with \mytool  in \autoref{sec:implementation}.
Our results are discussed in \autoref{sec:discussion}. Finally, our work is concluded in \autoref{sec:conclusion}.

%\todo{complete outline ... do we do some evaluation? can we measure sth on the device we built?}

%!TEX root = ../finder.tex

\section{Related Work}
\label{sec:related}

Heiland and Compton discovered and disclosed \num{12} vulnerabilities in \emph{TrackR Bravo}, \emph{iTrack Easy}, and \emph{Nut}~\cite{rapid7}.
Their analysis focused on the Bluetooth interface and forging finder identities to access data.
We assume that they used the same standard procedure to analyze all Bluetooth finders since the vulnerabilities found
are very similar.
%Most of these vulnerabilities are local.
%, such as user passwords stored unencrypted inside the app.
Their most outstanding findings are:
\ac{GPS} positions of \emph{TrackR Bravo} can be queried by finder identifier (CVE-2016-6540);
duplicate registration of the same \emph{iTrack Easy} finder allows to access the device's \ac{GPS} data (CVE-2016-6543); and \emph{Nut} transmits reusable session tokens without using \ac{TLS}, thereby allowing \ac{MITM} attacks on specific accounts (CVE-2016-6548).
The first and second attack require a valid device identifier, meaning that an attacker must either guess it or launch a targeted attack. The third attack's severity depends on the attacker's network position.
%---it is hard to perform for an individual, but easy for network administrators.
%
However, the security analysts did not find any \emph{Tile} issues despite analyzing it. Yet, we were able to uncover 
security issues and data leaks, detailed in \autoref{sssec:tile_sec}.
%We assume they did not invest the time to analyze communication keys retrieved from \emph{Tile}'s cloud infrastructure,
%which enabled further analysis and uncovering severe issues in \autoref{sssec:tile_sec}.
To the best of our knowledge, Heiland and Compton are the only ones who did security research on Bluetooth finders. No more vulnerabilities on the products we analyzed were listed in the \ac{CVE} database.
%, which is very surprising regarding the sheer amount of finder variants.

%More research has been done on sophisticated hardware that already receives \ac{GPS} information itself and has a direct Internet connection. For example, last year, an emergency device for older people has been shown to be exploitable~\cite{findus}. Similar to some Bluetooth finders we analyzed, they are manufactured in China and sold under various names. Since the vulnerable logic is in the hardware itself, the emergency devices need to be replaced---with more than \num{10000} devices affected in the UK alone.

Only recently, \emph{Apple} added the \emph{Find My} protocol on most of their devices~\cite{findmy}. As of now, \emph{Find My} only runs on powerful devices that have a regularly charged battery. All these devices are associated with a validated \emph{iCloud} account. This enables \emph{Apple} to run a complex encryption scheme. In comparison, the finders in this paper are powered by a button cell for more than a year, and \mytool does not enforce the privacy-invasive equivalent of an \emph{iCloud} account.

%\todo{this is a bit redundant, maybe move it to implementation}
%We design a new finder protocol \mytool and implement it, which makes existing implementations of Bluetooth finders another topic for related work. \acp{SDK} for hardware include code examples.
%A very simple application for Bluetooth often included in \acp{SDK} is a basic finder, which beeps on connection loss.
%Even development hardware in the correct size factor with this application exists, for example, the \emph{nRF51822 Bluetooth Smart Beacon Kit}~\cite{NRF51822-BEACON}.

%!TEX root = ../finder.tex

\section{Analysis of Existing Finders}
\label{sec:analysis}
In \autoref{ssec:features} we compare features of top-selling finders.
This is followed by a security analysis in \autoref{ssec:security}.
We conclude common features and issues in \autoref{ssec:analysis_results}.

\begin{table*}[hpt] 
    \caption{Finder features compared.}
    \footnotesize
    \begin{tabular*}{\linewidth}{ p{2.6cm} p{2cm} p{1.4cm} p{2.7cm} p{1.7cm} p{2.6cm} p{3cm}}
    \label{tab:finder_features}

        \textbf{\backslashbox[0pt][l]{Feature}{App}} & \textbf{Nut\newline Smart Tracker} & \textbf{Tile} & \textbf{musegear finder}, \newline \textbf{iTrackEasy} & \textbf{Cube Tracker} & \textbf{keeper} &  \textbf{iTracing},  \textbf{iSearching}, \newline \textbf{FindELFI} \\
        \hline
        \textbf{Device} & Nut Find3 & Tile Mate,\newline Tile Pro & musegear finder,\newline Pearl Callstel Key Finder & Cube Tracker & Gigaset g-tag  & Based on ST17H26 \\

        \hline
        \textbf{Declare lost} & $\checkmark$ & $\checkmark$ &  $\checkmark$ & $\checkmark$ & $\times$  & $\times$ \\
        \hline
        \textbf{Reports location \newline to server} & yes & yes & yes & yes & ?  & no \\
        \hline
        \textbf{Login} & Email/Phone+PW, \newline Social Login & Email+PW, \newline Social Login & Email+PW & Email+PW, \newline Social Login  & Email+PW \newline (\emph{Gigaset elements account}) & $\times$ \\

    \end{tabular*}
\end{table*}

\subsection{Feature Comparison}
\label{ssec:features}

In this section, we compare the features of a large selection of finders.
An overview is shown in \autoref{tab:finder_features}.
Despite being sold under different names, finders often share the same slightly customized
firmware and \ac{API}.
We perform an initial app analysis to avoid buying finders that are hardware or app duplicates by inspecting the app's code structure and \ac{API} calls.
We unpack and analyze the \emph{Android} apps by searching for strings and logic analysis based on \emph{jadx}~\cite{maddie}.

\subsubsection{Nut}
The \emph{Nut} finders support group search for lost devices and silent zones~\cite{nut-silent}.
Group search
means that the user can share the finder to friends with a QR code which can then help to locate the
finder~\cite{nut-groupshare}.
\emph{Nut} is ranked \emph{\#2} on \emph{Amazon} within the category \emph{GPS, Finders \& Accessories}.

The \emph{Android} and \emph{iOS} apps differ a lot.
Only the \emph{iOS} app has buttons for a lost finder search, but it never issued a request to the server in practice.
An \ac{API} call requesting lost finders exists in the dissected \emph{Android} app but
is never called. 
We assume some parts of the app were never implemented.
Nonetheless, both apps report all found devices and forward them to the \emph{Nut}
servers in the background. 
Locations are uploaded frequently---while the app is connected to the finder or when
it senses other disconnected finders.
Even though locations are sent to the server, the search feature is not available for users.

\subsubsection{Tile}
The leading finder on \emph{Amazon}, ranked \emph{\#1} in  \emph{GPS, Finders \& Accessories}, is \emph{Tile}.
While \emph{Nut} and \emph{Tile} features are similar, \emph{Tile} is more expensive and most features require a monthly subscription.
Basic \emph{Tile} features are to let the finder play a sound and to use finders to locate a smartphone.
Users can add multiple smartphones to their profile and also associate a limited subset of paired non-\emph{Tile} Bluetooth devices. \emph{Tile} comes with a crowd search feature and claims to have the largest finding community~\cite{tile-howto}.
Once a user account is associated to a \emph{Tile}, the account and the tracker are permanently bound. %This should be written passively, as one would also not write "his" account.
Losing account access bricks all associated \emph{Tiles}. 

The location is reported every few minutes to the servers.
In addition to the exact location, each click action in the app sends metadata to the server, such as the currently used Wi-Fi name and MAC address, which also allows inferring a location. Similar data is transferred every few minutes if no action is performed.

\subsubsection{musegear \& iTrackEasy}
Several finders with a similar casing design appear on the
first page of the \emph{Amazon} category \emph{GPS, Finders \& Accessories} under various
product names. One variant of this is the \emph{Pearl Callstel Key Finder}. This type of finder uses
the same \ac{API} hosted on different servers.
The user's location is reported to a server every \SI{30}{\minute} even in the absence of a finder, which is more privacy-violating than the \emph{Nut} implementation that only works with physically present finders. Also, the user's location is transferred when the user logs in and when a lost finder is reported.
When a user marks a finder as lost~\cite{musegear}, the server will report its location to the owner once it is found.

\subsubsection{Cube Tracker}
The \emph{Cube Tracker} is also amongst the most popular finders.
The \emph{Cube Pro} has twice the range as the basic \emph{Cube Tracker}.
Both finders have a crowd search, a replaceable battery, and a photo trigger function~\cite{cubetracker}. Also, they can play a sound on the finder and make the phone ring.
Similar to \emph{Tile}, the \emph{Cube Tracker} has an online sign-in that can be used to locate a finder.

\subsubsection{keeper}
We chose the \emph{Gigaset g-tag} for its different technologies and app codebase, although it is not a top-selling finder. 
The app requires a unique \emph{Gigaset elements account}.
Users can edit their profile with the app, but the account is not required for anything---except that it is enforced to have an account.
Locations are stored locally in a \emph{Realm}~\cite{realm} database and seem to be never transmitted to the server. The app's rating is poor because users expect to see their finder's location within their profile.

\subsubsection{iTracing, iSearching, FindELFI}
We selected the family of \emph{iTracing}, \emph{iSearching} and \emph{FindELFI} finders because they are one of the cheapest top-sellers on \emph{AliExpress}.
They look identical and also share the same hardware design. They are all based on the \emph{ST17H26} chip.
Due to their simple hardware design, there is no possibility to update the finder's firmware. The apps lack all cloud-based features.

%\FloatBarrier

\begin{table*}[htpb]
    \caption{Common finder analysis results.}
    \label{tab:finder_analysis}
    \footnotesize
\scalebox{0.9}{%
	\centering
    \begin{tabular*}{1.11\linewidth}{ @{\extracolsep{\fill}}  p{2.2cm} p{1.8cm} p{2.1cm} p{2.4cm} p{3.2cm} p{2.8cm} p{2.3cm}}
        \textbf{\backslashbox[0pt][l]{Analysis}{App}} & \textbf{Nut\newline Smart Tracker} & \textbf{Tile} & \textbf{musegear finder}, \newline \textbf{iTrackEasy} & \textbf{Cube Tracker} & \textbf{keeper} & \textbf{iTracing},  \textbf{iSearching}, \newline \textbf{FindELFI} \\
        \hline
        \textbf{TLS cert validation} & ignored & yes & pinning & pinning & yes & n/a  \\
        \hline
        %\textbf{Vulnerabilities} & \ac{API} traffic \ac{MITM}\newline Insecure share links\newline User data and location leak & MQTT wildcard subscription &  &  &  &   \\
        %\hline
        \textbf{API authentication} & no & AWS\newline credentials & Client\newline certificate & AWS\newline credentials  & Client \newline certificate & Basic \newline authentication  \\
        \hline
        \textbf{FW location} & Server\newline via HTTP & Amazon AWS\newline via HTTPS & n/a  & App & App & n/a \\
        \hline
        \textbf{FW encryption} & App & none & n/a  & none & Finder & n/a \\
        \hline
        \textbf{App obfuscation} & $\checkmark$ & $\times$ & $\times$ & $\checkmark$ & $\times$ & $\checkmark$  \\
        \hline
        \textbf{Package name} & com.nut. \newline blehunter & com. \newline thetileapp & com.antilost.finder, \newline com.antilost.app3 & com.blueskyhomesales.cube, \newline com.shenzhen.android.cube & com.gigaset.elements.\newline android.app.gtag2 & com.fb.antiloss,  com.lenzetech.antilost,  com.zoqin.findelfi  \\
        \hline
        \textbf{API endpoint} & \url{https://api.find.nutspace.com/},\newline \url{https://qa-find.nutspace.com/} & \url{https://production.tile-api.com},\newline \url{https://locations-prod.tile-api.com} & \url{https://api.musegear.net:8092/CommApiEx}, \newline \url{https://api.ieasytec.com:8092/CommApiEx} & AWS & \url{https://api.gigaset-elements.de/api/v1/}, \newline \url{https://im.gigaset-elements.de/identity/api/v1/} & - \\
		\hline
		\textbf{Individual issues}	& Leakage of all user and location data & MQTT data leakage and ring control & Static TLS keystore, \newline registration XSS & Outdated software, fake account registration, prototype pollution internal server error & --- & Duplicate apps, \newline usage statistics \\
		\hline
		\textbf{Disclosure} & 10/7/2018, unfixed & 5/2/2019 & 5/13/2019 & 1/28/2020 & --- & --- \\

    \end{tabular*}
} % end of adjustbox
\end{table*}

\subsection{Security Analysis}
\label{ssec:security}
We perform a technology and security analysis for all finders. Possible attack vectors strongly depend
on server-related features.
The general analysis follows the same categories.
In addition to these categories, we perform a device-specific analysis. This analysis includes app and cloud services.
An overview of the common analysis results is shown in \autoref{tab:finder_analysis}, and the individual
security issues are discussed in the corresponding subsections. 

The common analysis steps and categories are as follows.
Depending on the \emph{\ac{TLS} certificate validation} scheme, \ac{MITM} traffic analysis requires installing a root certificate to the smartphone, or a server certificate needs to be replaced within the app.
For \ac{MITM} traffic analysis, we use \emph{mitmproxy} and \emph{Burp Suite}~\cite{mitmproxy, burp}.
The category \emph{API authentication} summarizes how the app authenticates with the server. %summarizes what the smartphone app does to convince
% the server of its authenticity.
Typically, such \ac{API} authentication credentials can be exfiltrated from the app.
\emph{Firmware location and encryption} refers to firmware updates if the finder supports it.
Updates are forwarded by the app to the finder because it only communicates via Bluetooth.
The app caches the update in some location, ideally encrypted. In some cases, we were able to download firmware updates directly from the server or to decrypt the updates once they were downloaded by the app.
Ideally, the \ac{OTA} update keeps the firmware encrypted at all times until it is decrypted and verified on the finder.
If the app is \emph{obfuscated}, analyzing or changing its functions is more complicated. Non-obfuscated apps come with function names. All apps contained debug strings, which made de-obfuscation easier.
The \emph{package name} and class structure of \emph{Java} applications
as well as \emph{\ac{API} endpoints}, hint to a shared codebase between differently branded products.

\subsubsection{Nut}
The \emph{Nut} ecosystem implements a rich set of features but comes with the most security issues.

\paragraph{Traffic Interception}
The first step for analysis of the app is an \ac{MITM} attack on \ac{TLS} to
inspect the app's behavior.
%Apps communicating with only one server typically prevent
%this by pinning the server's certificate.
Instead of checking \ac{TLS} certificates, the app ignores any
\texttt{CertificateException} within the class \texttt{X509TrustManager}.
Thus, even an obvious \ac{MITM} attack with invalid certificates is possible
without installing new certificates to a victim's smartphone.
Anyone with access to the same network can sniff and manipulate \emph{Nut}
app traffic.
%We assume \emph{Nut} had issues with installing valid \ac{TLS} certificates on their
%servers while developing the app. Nobody removed the \ac{TLS} certificate check override
%in production since everything seems to be working correctly.

\paragraph{Firmware Leakage}
All \emph{Nut} firmware, including unreleased product series, can be retrieved from the server.
They are encrypted, but before installing an \ac{OTA} update to a finder
it is locally decrypted in the app with a static key.
%We were able to download the most recent firmware of a finder under development.
%The new generation of finders does not require an app since it comes
%with a \ac{GPS} module and a baseband modem. However, this makes the hardware
%more costly and opens new attack vectors~\cite{findus}. The new \ac{GPS}-based product is not yet available in the online shop, which means that not only the firmware is leaked but also \emph{Nut}'s production and marketing plans.

%\lstset{basicstyle=\normalsize\ttfamily}
\paragraph{Group Sharing}
%\todo{polish this paragraph by adding a simple graph of dependencies}
A group sharing link allows location information access to other \emph{Nut} users.
We are able to generate arbitrary group share links for known finders of other users
without using the app's QR code group share feature~\cite{nut-groupshare}.
Share links are supposed to be generated locally inside the app.
They contain the finder's \lstinline{deviceUUID}, an \lstinline{expirationTime},
and something internally called \mbox{\lstinline{hmac}.} This \lstinline{hmac} is not
what the name suggests since it is missing an encryption key. Instead, it is generated
only from known values:
\begin{center}
\vspace{-0.5em}
\begin{lstlisting}[basicstyle=\small\ttfamily]
  hmac=sha1(userUUID|deviceUUID|expirationTime)
\end{lstlisting}
\end{center}
%\lstset{basicstyle=\normalsize\ttfamily}

This share link is used to request a \lstinline{shareRecordUUID} from
the server. In the next step, a \lstinline{shareRecordUUID} can be used
to obtain a \lstinline{shareRecord}. This \lstinline{shareRecord} contains
the share's \lstinline{userUUID} as well as the \lstinline{userUUID} of
all users the finder is shared with.
Moreover, some privacy-concerning details such as the last known position of a
finder are included.

This results in two vulnerabilities.
First, an attacker can create share links for all devices with a known \lstinline{deviceUUID} and
\mbox{\lstinline{userUUID}.}
Second, an attacker can generate a new share link and invite themself even if the
previous share expired or was revoked by the owner since \lstinline{deviceUUID} and \lstinline{userUUID} leak with a share link.

\begin{figure}
\includegraphics[width=\columnwidth]{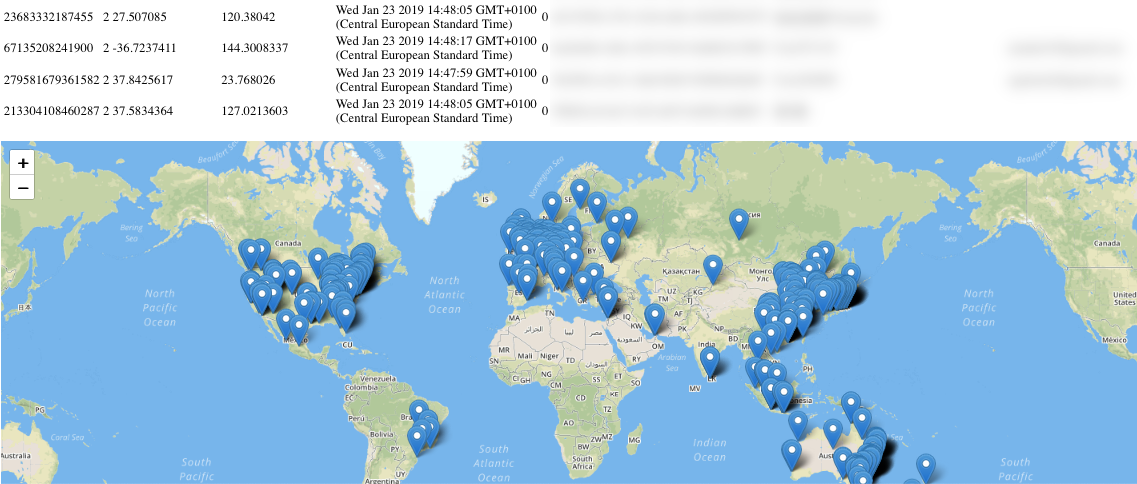}
\caption{Anonymized snapshot most recent \emph{Nut} reports.}
\label{fig:nut-lost}
\end{figure}

\paragraph{Retrieving All Lost Finder Reports}
All users of the \emph{Nut} app report lost finders.
Only lost finders appear during a Bluetooth device scan, connected finders are invisible.
Besides the lost finder's ID and \ac{GPS} position,
a report also contains the reporter's identity, including their mail address, phone number, and password hash.

An undocumented endpoint in the \ac{API} leaks all recently seen devices to the public, which represents a severe privacy
issue. The endpoint is contained in the app, but is never accessed during normal app operation.
\emph{Nut} did not close this issue more than a year after reporting it.
Thus, we do not reveal the actual \ac{API} call.
\autoref{fig:nut-lost} illustrates an anonymized example of the attack's impact. We limited the request to the server
to the last few lost finders. 
Even though the \ac{API} function requires the user's \lstinline{device_id} as an
argument, it is ignored.

%\end{lstlisting}

\paragraph{Responsible Disclosure}
We wrote \emph{Nut} an email on
October 7, 2018. On January 18, 2019 we
sent them a letter to their mailbox in California. We also
sent them a Twitter direct message on February 4, 2019.
Due to our unsuccessful attempts, we reported the issue to BSI \ac{CERT} on March 5, 2019 who then
forwarded it to US \ac{CERT} without further response.
Thus, BSI forwarded the information to \ac{CERT}/CC on April 30 2019.
We contacted \emph{Nut} again over Facebook on April 3, 2019 and called the number listed on their Facebook site but were not able to contact them.
We were also able to obtain the email address of one of the developers---we sent them an email in Chinese on April 5, 2019.
\ac{CERT}/CC recommended us to publicly request \acp{CVE} disclose our findings.
The missing certificate validation in the app was assigned CVE-2019-16252.
However, server-side issues in custom applications, as it is the case for the remaining issues, are not eligible for a \ac{CVE}.

\subsubsection{Tile}
\label{sssec:tile_sec}
Cloud features in the \emph{Tile} ecosystem are more secure than in the \emph{Nut} ecosystem.
The notification backend is based on \ac{MQTT} and properly using \ac{TLS} and authentication.
Yet, specific implementation details make their infrastructure vulnerable.

\paragraph{Firmware Leakage}
Firmware is available on the server in plaintext.
A valid link for the associated finder type can be extracted during the firmware update process.

\paragraph{Public Static Credentials for MQTT Server}

The app uses an \ac{MQTT} server to allow remotely ringing the phone via another device.
Credentials for \ac{MQTT} access are requested over a public \ac{API} endpoint.
This way, credentials are not stored inside the app---and could change regularly.
However, the \ac{MQTT} credentials returned by the \ac{API} are static and the same for all users.
Thus, there is insufficient authentication towards the \ac{MQTT} backend and no user separation.

\paragraph{Controlling Other Users' Phones}

Once connected to the \ac{MQTT} server, a request to ring any connected phone can be issued.
A valid request must contain a \lstinline{tileUUID}, but no further access control
is performed on requests. A \lstinline{tileUUID} can leak over various ways. For example, it can be requested from
any nearby \emph{Tile} via Bluetooth, and remote attackers can extract it with the attack described next.

\paragraph{User Data Leakage}

In \ac{MQTT}, data is exchanged by publishing it to topics to which clients can be subscribed.
Topic subscriptions allow for wildcards by default~\cite[p. 57]{mqtt}. \emph{Tile} blacklisted multi-level wildcards represented with \lstinline{#}, but did not include blacklist single-level wildcards with \lstinline{+}.
When subscribing to the wildcard topic \lstinline{+}, an attacker can receive all \ac{MQTT} messages, including messages meant for all \emph{Tiles} and system messages.
Specific messages that can be intercepted include regular status messages of the type \lstinline{CONTROL_STATUS_CHANGED} sent by the phones, which include the user's mail address, the user's \lstinline{UUID}, client \lstinline{UUID}, and \mbox{\lstinline{tileUUID}}. Combined with the previous knowledge, the \lstinline{tileUUID} can be used to ring the user's phone.
An attacker can also subscribe to the wildcard topic \lstinline{$sys/#}, the so-called sys topics. The broker responds with its uptime, program name, and version (\lstinline{Erlang MQTT Broker 2.2}), as well as real-time notifications of other clients connecting and disconnecting, including their client \lstinline{UUID}, user name, and IP address.

\paragraph{Responsible Disclosure}
We disclosed all issues with detailed explanations to \emph{Tile} on May 2, 2019.
They replied within one day and applied fixes in a timely manner.

\subsubsection{musegear \& iTrackEasy}
Attacks found in the \emph{musegear} finder family are rather weak. However, they indicate design flaws that might originate from insufficient security testing.
%Most likely
%server and app implementation did have some basic security testing, but some design
%flaws remain.

\paragraph{Keystore}
For customization, the app allows the vendor to configure different server locations.
Since servers use individual \ac{TLS} certificates, the app also has a local keystore
with trusted certificates. The password to protect this keystore is static and the same
in both apps, \emph{musegear}, and \emph{iTrackEasy}. This enabled us to modify the app
and run a \ac{MITM} attack for further analysis.

\paragraph{Registration Link}
After creating an account, the user receives an email with a registration confirmation link.
It is possible to include \ac{XSS} contents in the link to change the website's appearance
in the browser.

\paragraph{Responsible Disclosure}
We informed \emph{MS kajak7 UG} and \emph{KKM Company Limited}, the companies behind \emph{musegear} and \emph{iTrack Easy},
on May 13, 2019. \emph{musegear} reacted within less than an hour to our first contact attempt, and we discussed all findings.

%\todo{maybe write that both these attacks are quite harmless but show bad design}

\subsubsection{Cube Tracker}
While we could not find any severe data leakage within the \emph{Cube} ecosystem, there are a lot of 
minor security issues.
Their infrastructure is hosted on \ac{AWS}, and all traffic passes an \emph{Nginx} server.
Further services running in the backend are \emph{AngularJS}, \emph{ExpressJS}, \emph{Mongoose}, and \emph{MongoDB}.

\paragraph{Firmware Leakage}
The firmware is stored locally in the app. It is neither signed nor encrypted.

%\paragraph{Web Application Misconfiguration}
%The webserver is missing the \ac{HSTS} header, which is
%recommended by OWASP to always enforce usage of HTTPS instead of HTTP~\cite{owasp-hsts}.
%Moreover, the cookie set is missing the \texttt{SameSite} property.

\paragraph{Query Existing Users}
The login website displays different error messages for login failures versus non-existing users.

\paragraph{Fake Account Generation}
Instead of confirming an email on account generation, a user can call the \ac{API} endpoint
\lstinline{PUT /users/<id>/edit}. This allows the attacker to create an endless amount of accounts
without the mail overhead.

\paragraph{Profile Picture}
The profile picture can be set to an external URL. However, as far as we observed its usage, this picture
is only presented to the user themself.

\paragraph{Server-side JavaScript Prototype Pollution}
It is possible to alter \emph{JavaScript} values by sending malicious JSON payloads to \emph{Mongoose}.
For example, we were able to pass a payload that turns the \lstinline{User} type into a
generic \lstinline{Object} type. When \lstinline{User}-related methods are then called on the
\lstinline{Object}, this causes an internal error of type \lstinline{500} because the method is not found.
\todo[color=yellow!30]{\emph{Reviewer C: Which error?} 

explained it --jc}
Note that this happens only within that request, it does not escalate into the whole web service.
We cannot exclude that remote code execution becomes possible via this issue as we do not have access to
the server's implementation, but assume that this is very unlikely.
%However, we were only able to trigger errors with this payload and could not exploit it.

\paragraph{Outdated Software}
The \emph{Nginx} webserver indicates \lstinline{nginx/1.13.7} in the header, which is an outdated version with
multiple publicly known \acp{CVE}.
Also, the \emph{AngularJS} version 1.3.20 is out of support.

\paragraph{Responsible Disclosure}
We contacted \emph{Cube Tracker} on January 28, 2020, and also sent a second mail containing more details.
They promptly confirmed the reception, but communication did not continue later, likely due to COVID-19.

\subsubsection{keeper}
We could not find anything remarkable. The app does not have any connected functionality besides the account, which is never used except for logging in.

\subsubsection{iTracing, iSearching, FindELFI}

Even though the apps for finders based on the \emph{ST17H26} chip do not report locations to the server according to what we observed, their apps lack privacy and user-friendliness.

\paragraph{Usage Statistics}
During installation, the app connects once to a server to transmit usage statistics. The statistics only contain finder information
but no user account.

\paragraph{App Duplicates}
There seem to be around 15 copies of the original app in \emph{Android's} \emph{Google Play Store}. In the best case, it is just customization for different resellers.
However, it might be that some copies contain malware or leak location data, even though \emph{Google} regularly checks for malware in apps.

\subsection{Analysis Results}
\label{ssec:analysis_results}

%\paragraph{Common Architecture}
The hardware implementation of all Bluetooth finders is rather simple.
Finders themselves are not aware of locations or user data.
Logic to find lost items is implemented in the app and cloud.
A common architecture issue is that the cloud is retrieving \ac{GPS} data from
all users in plaintext. No matter if there is a vulnerability in the implementation
or not, users need to trust the cloud operator to keep their location information private.
Most apps continuously transmit finder locations to the cloud, regardless of their lost state.
In the \emph{Tile} and \emph{Cube} ecosystems, which feature a Web platform where users can view their device locations from anywhere, this is reasonable.
In contrast, \emph{Nut} finders are bound to a smartphone installation anyway.
%We assume the main reason for transmitting and storing locations in plaintext on servers is that it resembles the easiest way to implement this feature.
%Though, since finders are very cheap for coming with lifetime cloud support, collecting location information might also be on purpose.
Location traces are very privacy-sensitive as they reveal a lot about the user's habits---and, thus, could be sold for marketing purposes and similar.
%On the one hand, this could be the laziness of the developers. On the other hand,
%this is very suspicious and vendors might do this on purpose to obtain location traces.
%\todo{MH: gibt es keinen möglichen legitimen Grund? (insgesamt zu unwissenschaftlich, das einfach zu behaupten / vorzuwerfen) - z.b. multi-phone support!}
Our security analysis shows that even the leading vendors fail to protect location and user data
in their clouds from being extracted by external attackers.

We claim that existing Bluetooth finders are privacy-invasive by design.
In most situations, \ac{GPS} data could be stored locally inside the app.
Only in lost mode, reporters need to transmit location information to the owner---which
can be end-to-end encrypted. Minimizing private data transferred and stored in
plaintext reduces the risk of data leakage if there are security issues.

%!TEX root = ../finder.tex

\section{\NoCaseChange{PrivateFind}: An Open, Secure, and Anonymous Finder Solution}
\label{sec:implementation}

In the following, we describe \mytool, which features a similar architecture and hardware design as existing commercial products.
\mytool enables finder crowd search without leaking private data.
It prevents data leakage by design, as the server never sees any \ac{GPS} locations in plaintext.
Moreover, it enables anonymous usage of the crowd search ecosystem.
It aims at preventing tracking by both the server infrastructure as well as other users.
\mytool is publicly available on \url{https://github.com/seemoo-lab/privatefind}.

We define two setup variants with different security and privacy guarantees in \autoref{ssec:registration}.
An anonymous lost finder reporting system is introduced in \autoref{ssec:report}.
Reports are independent of the setup procedure and, thus, compatible with both
 variants.
We discuss protocol design decisions in \autoref{ssec:variants}.
The hardware used for our open-source prototype and the corresponding \emph{Android} app is shown in \autoref{ssec:hw}.

\begin{figure*}[ht]
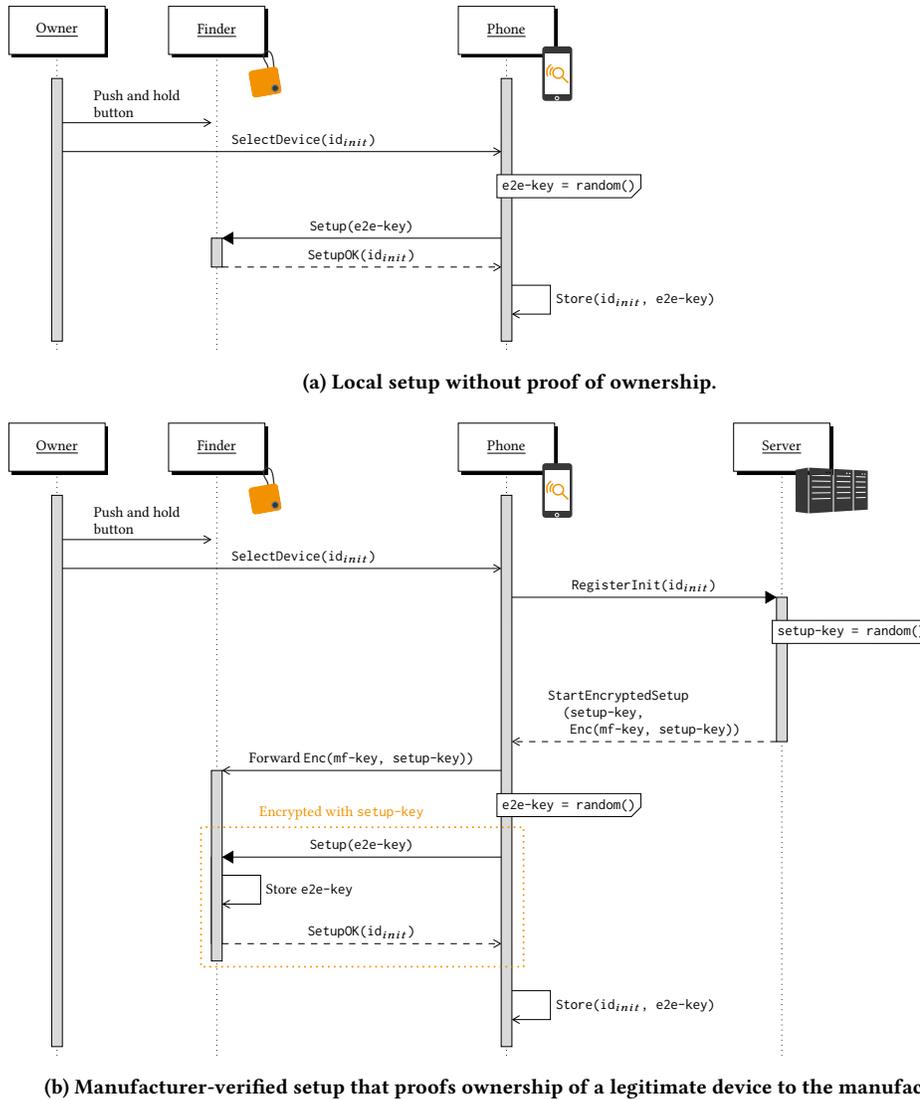


    \begin{subfigure}[b]{\textwidth}
\scalebox{0.8}{
  \footnotesize
  %\hspace*{-2.7cm}
  \mbox{
  \begin{sequencediagram}
    \newthread{U}{Owner}{}
    \newinst[1.3]{D}{Finder}{}
    \node[inner sep=0pt] (finder) at (8.3,-0.9)
    {\includegraphics[width=0.5cm,angle=10]{pics/finder.pdf}};
    \newthreadspace{P}{Phone}{4}
    \node[inner sep=0pt] (app) at (14.4,-0.9)
    {\includegraphics[width=0.5cm]{pics/smartphone.pdf}};
    \node[inner sep=0pt] (screenshot) at (14.4,-0.9)
    {\includegraphics[width=0.4cm]{pics/app.png}};
    \stepcounter{seqlevel} %increase level due to newline
    \messcall{U}{Push and hold\\ button}{D}
    \messcall{U}{\hspace*{3em}\texttt{SelectDevice(id$_{init}$)}}{P}
    
    \begin{note}{P}{\texttt{e2e-key = random()}}
    \end{note}
    \begin{call}{P}{\texttt{Setup(e2e-key)}}{D}{\texttt{SetupOK(id$_{init}$)}}
    \end{call}
    
    \begin{mess}{P}{\texttt{Store(id$_{init}$, e2e-key)}}{P}
    \end{mess}
  \end{sequencediagram}
  }}
  \caption{Local setup without proof of ownership.}
  \label{fig:register_anon}
  \vspace{1em} %spacing is too small, fix layout here

    \end{subfigure}    
    
    \begin{subfigure}[b]{\textwidth}
\scalebox{0.8}{
  \footnotesize
  %\hspace*{-2.7cm}
  \mbox{
  \begin{sequencediagram}
    \newthread{U}{Owner}{}
    \newinst[1.3]{D}{Finder}{}
    \node[inner sep=0pt] (finder) at (8.3,-0.9)
    {\includegraphics[width=0.5cm,angle=10]{pics/finder.pdf}};
    \newthreadspace{P}{Phone}{4}
    \node[inner sep=0pt] (app) at (14.4,-0.9)
    {\includegraphics[width=0.5cm]{pics/smartphone.pdf}};
    \node[inner sep=0pt] (screenshot) at (14.4,-0.9)
    {\includegraphics[width=0.4cm]{pics/app.png}};
    \newinst[3.7]{S}{Server}{}
    \node[inner sep=0pt] (server) at (20.1,-0.9)
    {\includegraphics[width=1.2cm]{pics/server.pdf}};
    \stepcounter{seqlevel} %increase level due to newline
    \messcall{U}{Push and hold\\ button}{D}
    \messcall{U}{\hspace*{3em}\texttt{SelectDevice(id$_{init}$)}}{P}
    %\messcall{P}{Initialize Bluetooth pairing}{D}
    %\messcall{D}{Confirm Bluetooth pairing}{P}
    %\messcall{P}{RegisterInit()}{S}
    %\messcall{S}{RegisterChallenge(setup-challenge)}{P}
    %\note{P}{e2e-key := random()}
    %\drawthread[2]{S}{D}
    \begin{call}{P}{\texttt{RegisterInit(id$_{init}$)}}{S}{\texttt{StartEncryptedSetup}\\ \hspace{1em}\texttt{(setup-key,} \\ \hspace{1.5em}\texttt{Enc(mf-key, setup-key))}}
    \begin{note}{S}{\texttt{setup-key = random()}}
    \end{note}
    \stepcounter{seqlevel} %increase level due to newline
    \stepcounter{seqlevel} %increase level due to newline
    \end{call}
    %\begin{messcall}{S}{\texttt{StartEncryptedSetup}\\ \hspace{1em}\texttt{(setup-key,} \\ \hspace{1.5em}\texttt{Enc(mf-key, setup-key))}}{P}
    \begin{messcall}{P}{Forward \texttt{Enc(mf-key, setup-key))}}{D}
    \draw[dotted,orange,thick] (7,-7.9) rectangle (13.7,-10.8);
    \draw (8.1, -7.6) node[orange,anchor=west] {Encrypted with \texttt{setup-key}};
    \begin{note}{P}{\texttt{e2e-key = random()}}
    \end{note}
    \begin{call}{P}{\texttt{Setup(e2e-key)}}{D}{\texttt{SetupOK(id$_{init}$)}}
    \begin{mess}{D}{Store \texttt{e2e-key}}{D}
    \end{mess}
    \end{call}
    \end{messcall} %
    %\end{messcall}
    \begin{mess}{P}{\texttt{Store(id$_{init}$, e2e-key)}}{P}
    \end{mess}
  \end{sequencediagram}
  }}
  \caption{Manufacturer-verified setup that proofs ownership of a legitimate device to the manufacturer.}
  \label{fig:register_mfg}
\end{subfigure}
\label{fig:reg}
\caption{Initial setup and registration variants.}
\end{figure*}

\subsection{Setup Procedure}
\label{ssec:registration}
The setup procedure establishes an end-to-end encryption key and identifiers, proves that the owner currently owns the finder, and comes in two variants.

\paragraph{End-to-end Encryption Key}
During the setup, the finder and the smartphone establish a pre-shared key \lstinline{e2e-key}.
The \lstinline{e2e-key} can be reset by running the setup procedure again.
Keys are individual per device; any leaked key will only compromise one finder. The \lstinline{e2e-key} is
symmetric due to the finder's hardware limitations.

After the setup, this key never leaves the finder and the smartphone. However, the finder can receive plaintext messages and encrypt
them with this key, such that only the smartphone can decrypt it. This mechanism is used to hide \ac{GPS} locations
of reports from the server by using the finder to encrypt reports about itself. Moreover, generating reports requires the physical presence of the respective finder.
Based on these reporting properties, the reporter can stay anonymous when sending reports to the server.

\paragraph{Initial and Randomized Identifier}
Moreover, the finder reveals its fixed identifier \lstinline{id}$_{init}$ during setup. This identifier never changes, even if the
finder is reset by repeating the setup procedure. The format of this identifier is implementation-specific, e.g., it
could be a 256\,bit random value. Both, the smartphone and the finder, use it in conjunction with the \lstinline{e2e-key},
to derive randomized identifiers in fixed time intervals: 

\mbox{\lstinline{id}$_{rand,n+1}$\lstinline{ = hmac(e2e-key, id}$_{rand,n}$\lstinline{)}}

\vspace{0.5em}

While \lstinline{id}$_{init}$ never leaves the finder and the smartphone (but is known to the server in the manufacturer-verified setup),
the randomized \lstinline{id}$_{rand}$ can be requested by nearby devices if the owner's smartphone lost the connection and if the finder
internally can confirm that it has not seen its owner since a while.
Thus, \mytool further increases privacy by only revealing \lstinline{id}$_{rand}$ if necessary. The finder does not appear in scan
results while it has a connection to its owner, and it can choose to refuse excessive identity requests, even on the randomized and regularly changing identifier.

\paragraph{Proving Ownership}
The setup mode that resets \lstinline{e2e-key} and leaks the unique identity \lstinline{id}$_{init}$ can only be performed
by the finder's owner.
To enter the setup mode, the owner pushes and holds the finder's button.
Only after pressing the button like this, the finder enters setup mode. Otherwise, it will not accept a reconfiguration.
Once the setup is finished, the setup mode must be reactivated again by pressing the button if needed.
We assume an attacker with physical access who wants to steal an item attached to a finder
could as well remove the finder's battery or shield it in tinfoil.
This is similar to the security assumptions made by other finder products.

\paragraph{Setup Variants}
While both setup variants establish an \lstinline{e2e-key}, exchange the \lstinline{id}$_{init}$, and ensure ownership
by physical access, they slightly differ.
The local variant in \autoref{sssec:register_anon} features a mode that enables compatibility between different finder manufacturers.
The manufac\-turer-verified variant in \autoref{sssec:register_mfg} enables the manufacturer to verify a finder's identity and provides further security to the Bluetooth communication.
% to restrict usage of
% their infrastructure by third-party products. % !!! no longer the case...
While we recommend using the local variant to improve privacy, manufacturers might
prefer the manufacturer-verified variant, as it is closer to the ecosystems provided by existing products. Both variants do not require the user
to register a personal account.

\subsubsection{Local Setup}
\label{sssec:register_anon}

The local setup, which is shown as a sequence diagram in \autoref{fig:register_anon}, ensures that the user has physical access to a finder.
Moreover, the setup exchanges \lstinline{e2e-key} and \lstinline{id}$_{init}$.
There is no third party involved in the local setup and there is no registration with a server.
Note that in this local variant \lstinline{id}$_{init}$ can also be reset to a random value during the setup,
as there are no external dependencies on it.

\subsubsection{Manufacturer-Verified Setup}
\label{sssec:register_mfg}

The manufactu\-rer-verified set\-up increases security
at the cost of privacy. It provides the same security properties as the local setup.
On top, it enables the manufacturer to validate that the finder is indeed one they created, and adds
manufacturer-verified encryption to the Bluetooth setup.
Even though finder validation makes the
server able to identify a finder during setup, messages containing precise \ac{GPS} location data will always
be end-to-end encrypted between the finder
and the user who registered a finder with their smartphone. Thus, also this protocol variant
ensures more privacy than any finder ecosystem we encountered in the wild.
The overall process is depicted in \autoref{fig:register_mfg}.

\paragraph{Manufacturing Requirements}
The manufacturer might want to verify their manufactured devices. %TODO property of only theirs in their ecosystem drops without token
To this end, the manufacturer installs an individual manufacturing key \lstinline{mf-key} on each finder,
which is associated with its identifier \lstinline{id}$_{init}$.
The \lstinline{mf-key} is the root of trust between server and finder.
Each finder has an individual \lstinline{mf-key}, which is
never transmitted in plaintext but can be identified by its \lstinline{id}$_{init}$.

%\FloatBarrier %TODO move this !!! -> fixed with setting figure placement to [h] and maybe the floats packet

\paragraph{Verification During Setup}
The finder notifies the server that \lstinline{id}$_{init}$ is now active.
\mytool does not enforce any account registration, but in case the manufacturer
would like to add this as a property, they could add a \lstinline{mf-key}-based
setup challenge in this step. Note that, depending on the remaining implementation,
this would also add reporter identities to the encrypted location reports.

\paragraph{Encrypting Bluetooth Communication}
We assume communication between app and server is not compromised since it can be protected with \ac{TLS}~\cite{RFC8446}. The server certificate must be validated and ideally is pinned~\cite{DBLP:conf/ndss/SounthirarajSGLK14}. Encryption methods employed by \ac{TLS} are not feasible on a low-cost finder device. Hence, the server generates a temporary random \mbox{\lstinline{setup-key},} which is used to encrypt the setup procedure. The communication during the setup, encrypted with \lstinline{setup-key}, is marked with an orange dotted box in \autoref{fig:register_mfg}. The \lstinline{setup-key} is transferred to the app in two formats: plaintext and encrypted.
The plaintext \lstinline{setup-key} is for the app; the app does not have the \lstinline{mf-key} to decrypt the encrypted one.
The app forwards the encrypted \lstinline{setup-key} to the finder via Bluetooth. Only a legitimate finder can decrypt it with its \lstinline{mf-key}.

This possibility of securing the registration procedure with the \lstinline{mf-key} improves security.
We do not consider Bluetooth encryption to be sufficient. Since the finder neither has a display nor a keyboard, \emph{Just Works} pairing is applied. Assuming \ac{BLE} 4.0 or 4.1
on at least one of the involved devices, this mode is susceptible to passive \ac{MITM} attacks due to weak encryption methods~\cite[p. 277]{bt52}. Even in \ac{BLE} 4.2 and higher, \emph{Just Works} can be actively eavesdropped, as stated in the Bluetooth 5.2 specification~\cite[p. 274]{bt52}.
Using the \lstinline{mf-key} as an additional root of trust mitigates this \ac{MITM} risk.

\begin{figure*}[!bp]
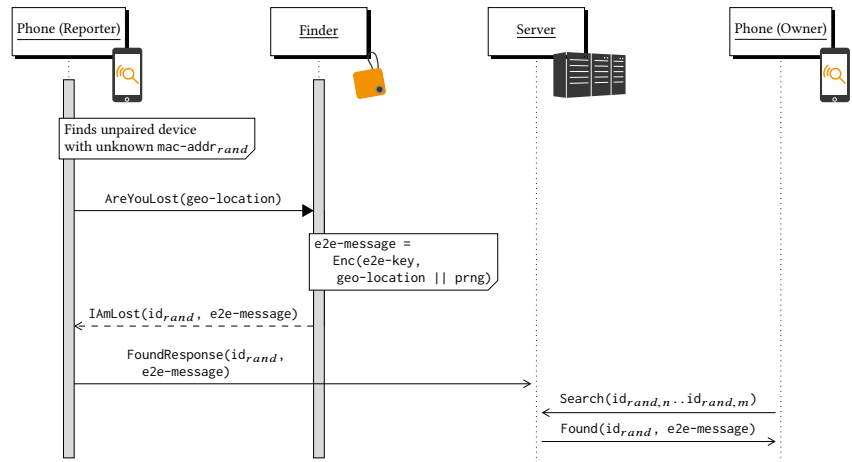

\scalebox{0.8}{
  \footnotesize
  \hspace*{-3cm}\mbox{
  \begin{sequencediagram}
    \newthread{F}{Phone (Reporter)}{}
    \node[inner sep=0pt] (app) at (5.4,-0.9)
    {\includegraphics[width=0.5cm]{pics/smartphone.pdf}};
    \node[inner sep=0pt] (screenshot) at (5.4,-0.9)
    {\includegraphics[width=0.4cm]{pics/app.png}};
    \newthread{D}{Finder}{}
    \node[inner sep=0pt] (finder) at (10.4,-0.9)
    {\includegraphics[width=0.5cm,angle=10]{pics/finder.pdf}};
    \newinst[2.5]{S}{Server}{}
    \node[inner sep=0pt] (server) at (15,-0.9)
    {\includegraphics[width=1.2cm]{pics/server.pdf}};
    \newinst[3]{O}{Phone (Owner)}{}
    \node[inner sep=0pt] (app) at (20.1,-0.9)
    {\includegraphics[width=0.5cm]{pics/smartphone.pdf}};
    \node[inner sep=0pt] (screenshot) at (20.1,-0.9)
    {\includegraphics[width=0.4cm]{pics/app.png}};
    \stepcounter{seqlevel} %increase level for icon distance
    \note{F}{Finds unpaired device \\ with unknown \texttt{mac-addr$_{rand}$}}
    \stepcounter{seqlevel} %increase level due to newline
    \begin{call}{F}{\texttt{AreYouLost(geo-location)}}{D}{\texttt{IAmLost(id$_{rand}$, e2e-message)}}
    \begin{note}{D}{\texttt{e2e-message =} \\ \hspace{1em} \texttt{Enc(e2e-key,} \\
    \hspace{1em}\texttt{ geo-location || prng)}}
    \stepcounter{seqlevel} %increase level due to newline
    \end{note}
    \end{call}
    \stepcounter{seqlevel}
    \messcall{F}{\mbox{\texttt{FoundResponse(id$_{rand}$,}\hspace{13em}}\\ \hspace{1em}\texttt{e2e-message)}}{S}
    \messcall{O}{\texttt{Search(id$_{rand,n}$..id$_{rand,m}$)}}{S}
    \messcall{S}{\texttt{Found(id$_{rand}$, e2e-message)}}{O}
  \end{sequencediagram}
  }
  }
  \caption{Anonymous report of witnessed device.}
  \label{fig:lost}
\end{figure*}

%\FloatBarrier

\subsection{Privately Reporting Lost Finders}
\label{ssec:report}

Based on the initial setup information, finders can
be located, as shown in the sequence diagram in \autoref{fig:lost}.
This mechanism does not leak the lost finder's location to the
server, and the reporter's identity is not revealed.

A lost finder can be found by anyone because it will be discoverable in Bluetooth scanning if it is not connected to the owner's smartphone. The reporter can send a message to a lost finder that contains the current \ac{GPS} position. The finder answers
this with an encrypted message that can only be decrypted by the owner with \lstinline{e2e-key}.
The finder only answers this message if it has not seen the owner for a few minutes to ensure that no unnecessary reports are generated. Finder and owner can use a pseudo-random sequence, if available on the finder, to prevent a replay of old locations. The current \mytool implementation prevents this with a counter and an authenticated encryption mode.
The owner can look up reports by its calculated recent list of \lstinline{id}$_{rand}$. %, or, as in the current \mytool implementation, by its Bluetooth MAC address.

\subsection{Implementation Variants}
\label{ssec:variants}

In the following, we discuss implementation variants and point out the current state of the \mytool implementation.

\paragraph{Identity Randomization}
The finder should randomize its MAC address as \lstinline{mac_addr}$_{rand}$ to prevent tracking by nearby devices.
Thus, the finder changes its address in regular intervals---otherwise, the randomization of the finder's identifier \lstinline{id}$_{rand}$ could be bypassed. This MAC address randomization feature is already
included in the Bluetooth specification to ensure privacy for BLE~\cite[p. 3064ff]{bt52}.
Usually, this address changes every \SI{15}{\minute}, but the interval can be lowered to increase privacy.
The interval of the MAC address change and \lstinline{id}$_{rand}$ update should be synchronized to prevent tracking via asynchronous changes.

Note that the current \mytool code release is not using MAC address randomization, because it requires
MAC addresses for a simplified, non-randomized finder identification during location reports.

\paragraph{Report Delivery Modes}
\mytool uses \emph{direct delivery} of reports to an owner.
For direct delivery, the owner messages the server with the currently valid \lstinline{id}$_{rand}$ or a sequence of recently valid identities.
%While this leaks the current identity of the owner, the addresses cannot be linked without the \ac{IRK}, and thus, no future or past addresses are known.
%In our case, the server needs to associate finder identities \lstinline{id} to smartphone access tokens \lstinline{token}. \todo{we did not write that above! private variant would be the other version...}
%It is important to note that the access token is not a security feature for the user, as it does not help with message decryption, but just a message filter. Nonetheless, access tokens prevent the server infrastructure from being misused as publish-subscribe service by third parties.

An alternative approach would be \emph{broadcast delivery}. Since reports are encrypted, locations would not leak.
This form of broadcast delivery hides the fact which finder was lost and found at which point in time from the server. However, someone who lost an item needs to observe all reports, which causes a lot of traffic and leaks recently valid identities.
%Another reason to decide against this approach is that no server-side validation is possible, such as blacklisting reports of fake devices.

In the current \mytool implementation, a reporter is not getting any feedback. Reporters cannot verify if a finder exists. This increases security against attackers trying to leak valid identities.

\paragraph{Verifying Reporters and Reports}
\mytool does not verify a reporter's identity on the server-side to increase privacy.
Owners can check locally if a report was indeed created by their lost finder if it
decrypts correctly and contains a valid number of a pseudo-random sequence.
Moreover, anonymous location reports can be displayed differently inside the app than locations observed by the user. This helps the user to check themself if the reported location seems legit.

The server could verify if the reported finder is real by sending a challenge similar to the one in the setup. The challenge can again make use of the \lstinline{mf-key} shared between server and finder, but with the reporter as a relay.
Such a challenge would make it impossible to replay reports already on the server-side. A disadvantage of this approach is an increased load on the server for a property the owner can also verify locally.
Moreover, it would require the server to ask for \lstinline{id}$_{init}$ to look up the according \lstinline{mf-key}, which would deanonymize reports.
%The server is still able to blacklist duplicate messages or spamming reporters.

The server could also restrict usage of their ecosystem to those users who legitimately purchased a finder.
For this, the manufacturer-verified setup needs to be extended as follows.
The server can send a challenge bound to a finder's \lstinline{mf-key} that can only be answered by a legitimate finder.
If this challenge is answered correctly, the server can either confirm a user's account registration with this, or
issue an account-less access token. Later, when users report lost finders or search for them, these actions can
be bound to that account or token. While the \ac{GPS} locations in such an approach remain encrypted, this
always reveals the user's or finder's identity. However, even this extension would still be more privacy-preserving
than existing approaches that all share \ac{GPS} locations to the server.
The current \mytool code release has an option for such a token.

\paragraph{Deciding if a Finder is Lost}

A finder that lost the connection to the owner's smartphone appears in Bluetooth scan results of other smartphones. This might create a lot of reports in crowded places with bad connection quality.
Hence, the reporter asks the finder if it has been disconnected for a while, instead of reporting it immediately.
Yet, this still imposes a privacy issue. There is a possibility that an owner lost the connection but still knows the item's location, e.g., if the smartphone battery is empty. The report includes the reporter's IP address, which can be used to make a raw guess on the finder's geolocation.
%The fact that the finder's MAC address is randomized further obfuscates the association between the finder's identity and it's geolocation. Note that IP address information is sent along with any information to the server, %but a privacy-concerned user can anonymize her IP address during setup.
%\todo{@Max Can we enhance the app by an option that the tracker always says it is not lost? i.e. for items that typically do not move}

Our \mytool implementation enables users to opt-out from receiving reports by setting a flag in the finder that disables the generation of reports. As long as this flag is set, the finder will never answer to an \lstinline{AreYouLost} message.

The server can keep track of owners who report their finder as lost and propagate this information to reporters.
Even though this enables all users to see which finders are currently lost, the public information is not associated with an IP address and the
included \lstinline{id}$_{rand}$ is randomized.

\paragraph{Metadata on the Server}
Users still need to trust the server in some means.
The server can estimate the geolocation by IP addresses.
Message timing also allows it to guess whether reporter and owner
are nearby and know each other.
However, in \mytool this information is only transferred if the finder
confirms to be lost.
%This is nothing that happens automatically in the background all the time in the app---in contrast to the default behavior in most commercial products.
In contrast to the default behavior in most commercial products, the amount of reports is minimized and additionally anonymized.
In general, users being concerned about geolocations leaking by IP addresses should use anonymization techniques such as Tor~\cite{tor}.
Users could be selfish and not send reports for anything they found but still profit from other reporters if they are concerned about leaking information when reporting.

\paragraph{Identity Export}
A user might use a finder with multiple phones or backup the finder association in case of phone loss, the \lstinline{e2e-key} could be exported with a QR code.
Without export, only one smartphone at a time is supported. To replace the smartphone, the finder must run through the setup procedure again, which resets the \lstinline{e2e-key}.

\subsection{PrivateFind Implementation}
\label{ssec:hw}

In the following section, we detail how we realized \mytool in hardware, firmware, and as an \emph{Android} app.

\paragraph{Hardware Platform}

\begin{figure}[!b]
  \centering
  \begin{subfigure}{.5\columnwidth}
    \centering
    \vspace{6em} %move to app bottom level
    \includegraphics[width=0.98\columnwidth]{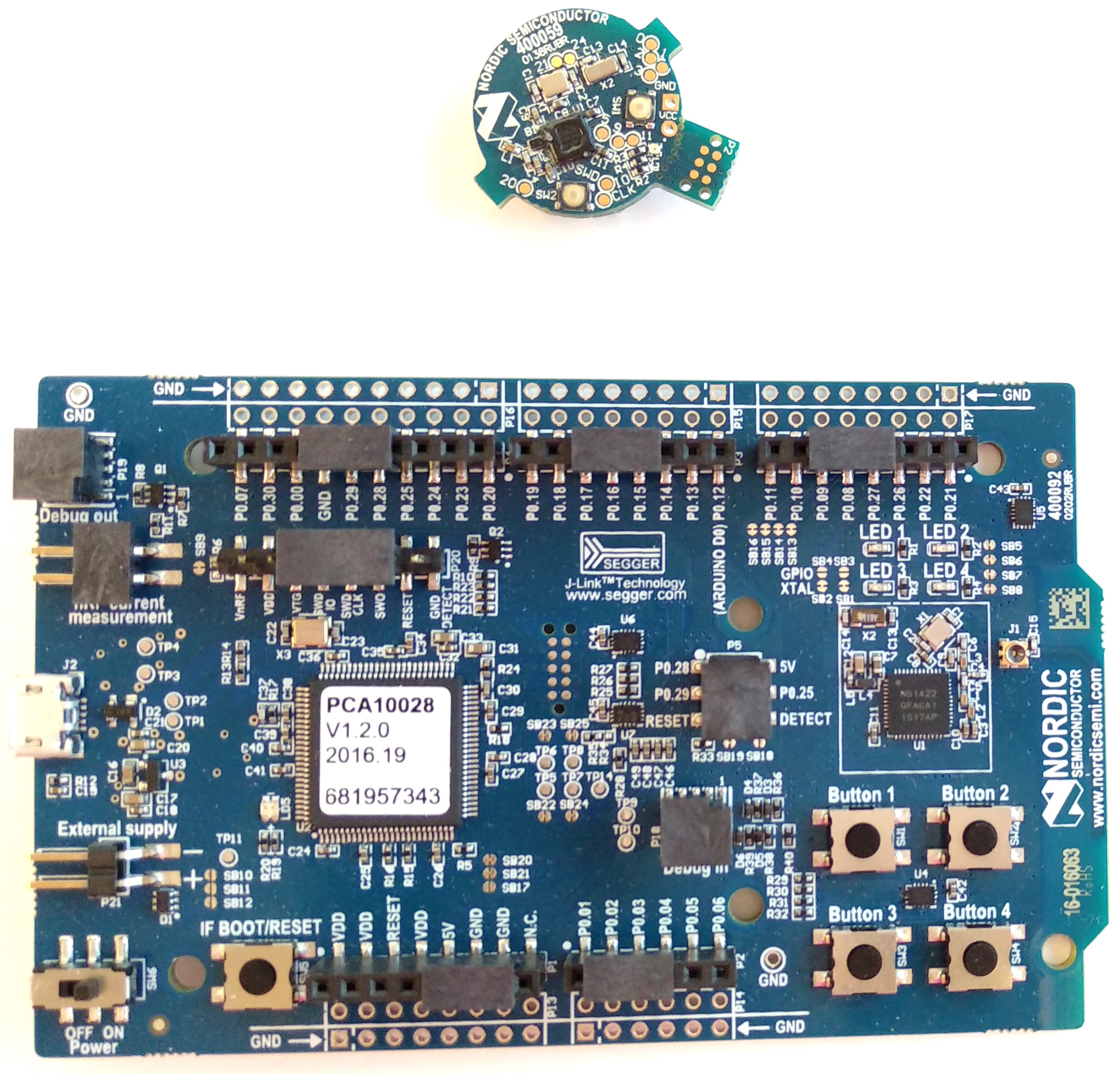}
    \caption{The two nRF platforms: \emph{nRF51-DK} (bottom) and \emph{nRF51822} (top).}
    \label{fig:nrf-devboards}
  \end{subfigure}%
  \begin{subfigure}{.5\columnwidth}
    \centering

    \scalebox{0.8}{

      \begin{tikzpicture}[minimum height=0.55cm, scale=0.8, every node/.style={scale=0.8}, node distance=0.7cm]
    
        \node[inner sep=0pt] (app) at (0,-0.7)
        {\includegraphics[width=5cm]{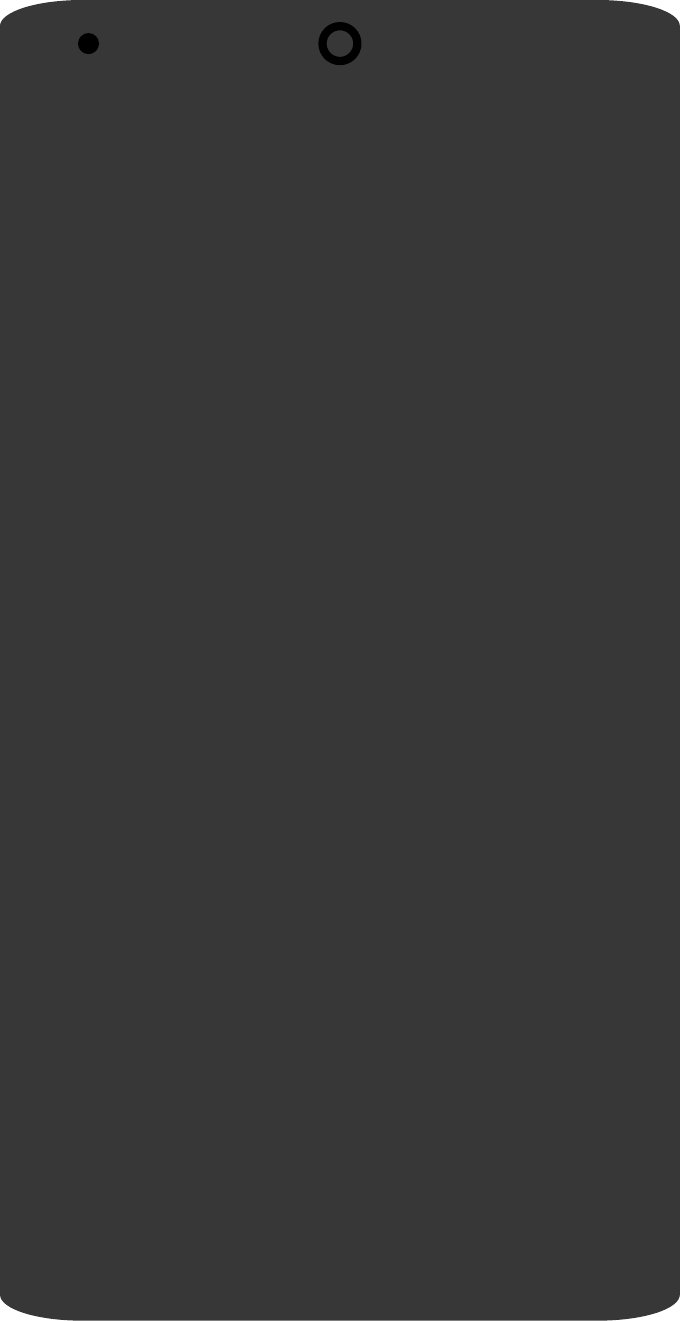}};
        \node[inner sep=0pt] (screenshot) at (0,-0.5)
        {\includegraphics[width=4.5cm]{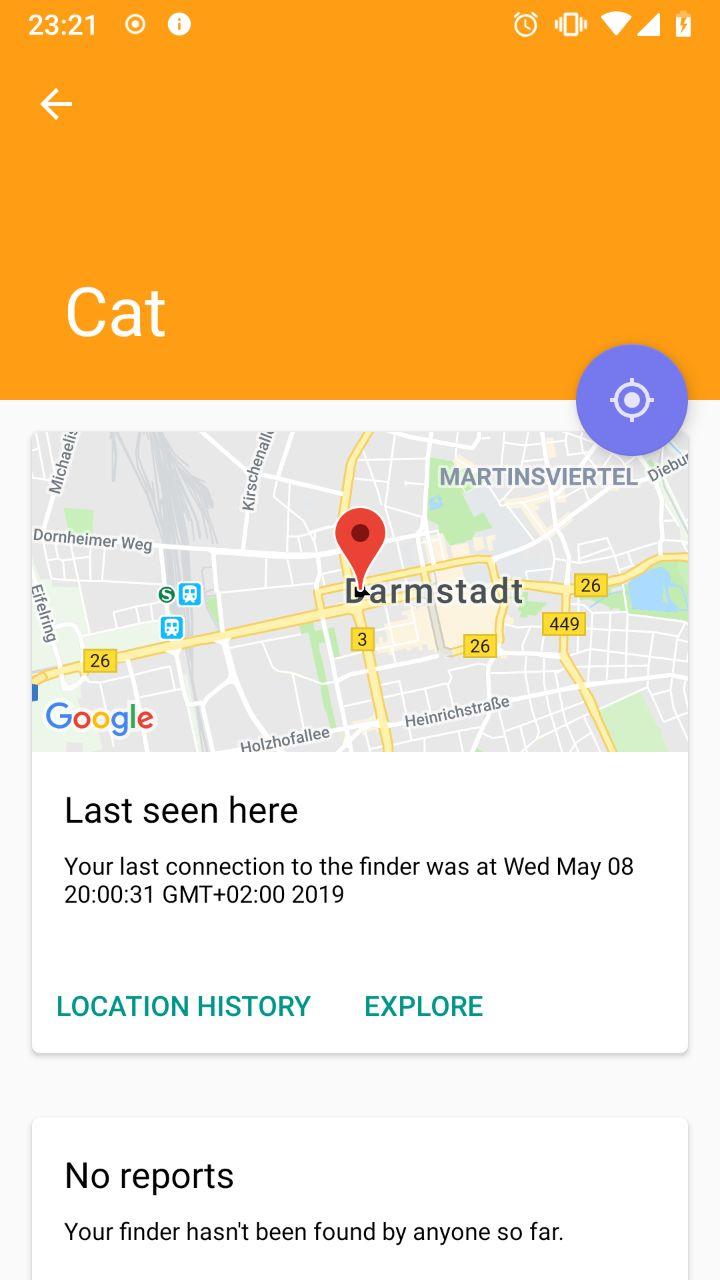}}; %we also have screenshot_berlin
    
      \end{tikzpicture}
    }
    \caption{Mobile app.}
    \label{fig:privatefindapp}

  \end{subfigure}
  \caption{\mytool implementation.}
  \label{fig:privatefindgeneral}
  \end{figure}

After disassembling common Bluetooth finders, we became aware of the \emph{nRF51822 Bluetooth Smart Beacon Kit}~\cite{NRF51822-BEACON}. Its diameter is as small as \SI{20}{\milli\meter}, and it comes with all the important features.
During development, we used the \emph{Bluetooth Low Energy Development Kit for the nRF51 Series}~\cite{NRF51-DK}. It is based on the same chip but meant for development, which means the board comes with additional input and output possibilities and is easier to flash---limitations for firmware running on the chip stay similar.

The development platform already comes with a basic finder example, which triggers an alarm on Bluetooth connection loss.
Only the \mytool setup, registration, and report procedures had to be implemented.
\todo[color=orange!30]{\emph{Reviewer D: What about the registration procedure?}

That's what ''setup'' meant but made it more clear. --jc}

\paragraph{Hardware Optimization}
The development platform already offers some encryption methods, such as \ac{AES}.
Encryption in the lost finder reporting is \ac{AEAD}, e.g.,
AES-CTR with a random nonce for the associated data and SHA256-HMAC for authentication.
Depending on the hardware platform, different encryption methods can be used.

\paragraph{Android Application}
The app runs in the background and maintains a Bluetooth connection with the finder established. This is implemented using the \emph{Android} \ac{BLE} library by \emph{Nordic Semiconductor}~\cite{nordic-ble}.
The last known position of a finder is saved locally and can be displayed as shown in \autoref{fig:privatefindapp}.
Once the connection is lost, it plays a sound on the smartphone or on the finder, according to the user's preferences.
Moreover, the app searches for other lost finders in the background and reports these.
To preserve battery on the smartphone, we use the hardware offloading of the Bluetooth controller if supported.

%!TEX root = ../finder.tex

\section{Discussion}
\label{sec:discussion}

\paragraph{Crowd Search}
Distributed item search requires a large user base with the app installed and running. For example, finders of three different brands were marked as lost, and the crowd search feature was not able to find them near a busy train station in Germany with approximately \num{250000} people passing by in 2017~\cite{ct-finder}. A lost item can only be found within communities that commonly use finders.
This problem gets worse with the diverse finder market since reporting systems are not compatible at all. We observed that online shop ratings of the same finders differ a lot, depending on the platform and region where they are sold, and assume this is due to the distinct communities using these finders.

\paragraph{Privacy Policies}
Besides the technical problems we uncovered in common Bluetooth finders, their basic concept is already privacy-invasive depending on how data is handled exactly. Thus, we did a check on the privacy policies
of finders with online functionality.
\begin{description}
  \item[Nut] There are two different privacy policies for \emph{Nut}. One is publicly available on their website but only refers to data their webserver collects while browsing their sites, and another one that is displayed when installing the app. We consider the policy in the app to be more relevant. Yet, the app privacy policy of \emph{Nut} is surprisingly short and written in very vague terms. The policy does not even state that \emph{Nut} is collecting location data. We assume that such a policy is not acceptable under the \ac{GDPR}.
  \item[Tile] Two privacy policies were in place at the same time before our report. When a user tried to create an account in the app, a privacy policy from 2016 was shown. However, on the website of \emph{Tile}, a new policy from 2018 is shown~\cite{tile-privacy}. The policy of the app states that the location of the user is transmitted periodically. It also states that \emph{Tile} may collect information for statistical purposes that cannot be traced back to an individual user anymore. In general, we consider the privacy policy of \emph{Tile} to be well-written and understandable.
  \item[musegear] The privacy policy by the \emph{musegear} app \todo[color=yellow!30]{\emph{Reviewer C: subjective formulation} fixed that --jc} looks very similar to \emph{Tile}. Their privacy policy states that location data should be regularly saved on the device, but does not mention the regular transmission of this data. Also, their policy is hard to read due to formatting mistakes.
  \item[CubeTracker] The privacy policy of \emph{Cube Tracker} is surprisingly well-written, but still contains a placeholder where the tax identifier of the company should be inserted. It allows \emph{Cube Tracker} to share data with third parties and also to update the policy, and the user will be notified about significant changes. Also, it does not differentiate which data is collected by the app locally and which data is shared with the cloud service and third parties. 
%  \emph{Cube Tracker} describes their attitude towards security as: \emph{``Cube Tracker has implemented various measures to enhance the protection of personal information and mitigate the risks of theft, damage, loss of information, or unauthorized access, use, disclosure, or alteration. {[...]} Please note that while we strive to protect personal information from risks, we do not provide any guarantee or warranty of information security.''}
  %\item[keeper] The privacy policy of keeper is short and very well-written. Since most connected features are missing, we think that it accurately reflects what the device and the app are doing.
\end{description}
To summarize, the privacy policies of all tested Bluetooth finders with connected features do not accurately reflect what the finders and corresponding apps are doing. Therefore, we would like to encourage the manufacturers to update their privacy policies. A general issue is that the privacy policies in the apps differ from the ones on the websites, and the policies are subject to change. This makes the privacy policies very opaque in addition to missing details on how data is processed and stored.

\section{Conclusion}
\label{sec:conclusion}

We conducted a comprehensive analysis of the most popular Bluetooth finders currently on the market and analyzed their security and privacy. 
None of the market-leading products is designed in a privacy-friendly way, and several of them have serious security flaws on multiple levels:

\begin{itemize}
  \item All products tested were designed and implemented without a focus on privacy.
  \item Some of the tested products forced the user to create an account for a cloud service not immediately required for using the product. Other products automatically reported the user's location to a cloud service without an observable reason.
  \item Some of the products had minor security vulnerabilities, such as insufficient protection of the \ac{API} against unauthorized communication. %Not exactly sure what this means, maybe drop "parties"?
   Furthermore, some products omitted proper \ac{TLS} certificate validation of the backend services.
  \item A few products had serious security vulnerabilities in their corresponding backend that enabled attackers to obtain access to private data of other users. %This is alread pointed out by "private" data. they are not supposed to access.
  \item One vendor ignored our reports about those weaknesses for more than a year, despite multiple attempts to get in touch with them over several communication channels. Until today, the corresponding cloud service leaks private customer information that can be accessed easily.
\end{itemize}

Some implementations are highly suspicious, and we cannot rule out that this data may be collected for other purposes. Many of those security and privacy problems can be fixed easily in the app and the corresponding backend services, for example by not sending any unnecessary data to the cloud service. We decided on a more thorough approach and designed and implemented the privacy-friendly Bluetooth finder \mytool that prevents location leakage to the cloud service. Users are able to report a lost and found Bluetooth finder anonymously, and we believe our system achieves a more advanced security and privacy standard than any commercial system we analyzed.
Our system provides the same features as most commercial products and runs on the same or comparable hardware. This shows that privacy-friendly and secure Bluetooth finders can be built without increasing expenses for the hardware and without a loss of features for the user.

\begin{acks}
We thank Max Maass and Nils Ole Tippenhauer for the discussion about the \mytool protocol, Vanessa Hahn for the graphics design,
and Matthias Hollick for his support.

This work has been funded by the German Federal
Ministry of Education and Research and the Hessen State Ministry for
Higher Education, Research and the Arts within their joint support of
the National Research Center for Applied Cybersecurity ATHENE.
\end{acks}

\bibliographystyle{ACM-Reference-Format}
\bibliography{bibfile}

\end{document}